\documentclass[11pt,letter]{article}
\usepackage{amsmath}
\usepackage{amsfonts}
\usepackage{amssymb}
\usepackage{graphicx}
\usepackage{enumerate}
\usepackage[ruled]{algorithm2e}
\usepackage{tabularx,booktabs}

\newtheorem{thm}{Theorem}[section]
\newtheorem{prop}[thm]{Proposition}

\newtheorem{lemma}[thm]{Lemma}

\newcommand{\QED}{\hfill$\square$}

\title {
    \bf {On vertex covers and matching number \\ of trapezoid graphs}
}

\author
{
{\large \sc Aleksandar Ili\' c \footnotemark[3]} \\
{\em \normalsize Faculty of Sciences and Mathematics, Vi\v segradska 33, 18000 Ni\v s, Serbia} \\
{\normalsize e-mail: { \tt aleksandari@gmail.com }}
\and
{\large \sc Andreja Ili\' c} \\
{\em \normalsize Faculty of Sciences and Mathematics, Vi\v segradska 33, 18000 Ni\v s, Serbia} \\
{\normalsize e-mail: { \tt andrejko.ilic@gmail.com }} }

\begin{document}

\maketitle

\begin{abstract}
The intersection graph of a collection of trapezoids with corner points lying on two parallel lines
is called a trapezoid graph. Using binary indexed tree data structure, we improve algorithms for
calculating the size and the number of minimum vertex covers (or independent sets), as well as the
total number of vertex covers, and reduce the time complexity from $O (n^2)$ to $O (n \log n)$,
where $n$ is the number of trapezoids. Furthermore, we present the family of counterexamples for
recently proposed algorithm with time complexity $O (n^2)$ for calculating the maximum cardinality
matching in trapezoid graphs.
\end{abstract}

 \vspace{0.2cm}

 \vspace{0.2cm}

\footnotetext[3] { Corresponding author. }

\section{Introduction}

A trapezoid diagram consists of two horizontal lines and a set of trapezoids with corner points
lying on these two lines. A graph $G = (V, E)$ is a trapezoid graph when a trapezoid diagram exists
with trapezoid set $T$, such that each vertex $i \in V$ corresponds to a trapezoid $T (i)$ and an
edge exists $(i, j) \in E$ if and only if trapezoids $T (i)$ and $T (j)$ intersect within the
trapezoid diagram. A trapezoid $T (i)$ between these lines has four corner points $a(i)$, $b(i)$,
$c(i)$ and $d(i)$ -- which represent the upper left, upper right, lower left and lower right corner
points of trapezoid~$i$, respectively. No two trapezoids share a common endpoint.

Ma and Spinrad \cite{MaSp94} showed that trapezoid graphs can be recognized in $O(n^2)$ time, while
Mertzios and Corneil \cite{MeCo09} designed structural trapezoid recognition algorithm based on the
vertex splitting method in $O (n (m + n))$ time, which is easier for implementation. Here $n$
stands for the number of vertices and $m$ for the number of edges. Trapezoid graphs are perfect,
subclass of cocomparability graphs and properly contain both interval graphs and permutation
graphs~\cite{Li07}.

Trapezoid graphs were first investigated by Corneil and Kamula \cite{CoKa87}. These graphs and
their generalizations were applied in various fields, including modeling channel routing problems
in VLSI design \cite{DaGoPi88} and identifying the optimal chain of non-overlapping fragments in
bioinformatics \cite{AbOh05}. Many common graph problems, such as minimum connected dominating sets
\cite{TsLiHs07}, all-pair shortest paths \cite{MoPaPa02}, maximum weighted cliques \cite{BePaPa02},
all cut vertices \cite{HoPaPa04}, chromatic number and clique cover \cite{FeMuWe97}, all hinge
vertices \cite{BePaPa03}, minimum vertex covers \cite{LiCh09} in trapezoid graphs, can be solved in
polynomial time. For other related problems see \cite{ChCo96,CrGa10,Li94,Li95}.

Let $G = (V, E)$ be a simple undirected graph with $|V| = n$ and $|E| = m$. A subset $C \subseteq
V$ is called a vertex cover (VC for short) of $G$ if and only if every edge in $E$ has at least one
endpoint in $C$. A vertex cover $C$ is called a minimal vertex cover if and only if no proper
subset of C is also a vertex cover. The size of a vertex cover is the number of vertices that it
contains. A vertex cover with minimum size is called a minimum vertex cover. Notably, a minimum VC
is always a minimal VC, but a minimal VC may not be a minimum VC.

Given a graph $G$ and a fixed parameter $k$, determining whether $G$ has a vertex cover of at most
$k$ vertices is one of the best-known NP complete problems \cite{Ka72}. Many recent studies have
focused on developing faster algorithms for the VC problem in special classes of graphs. The number
of vertex covers in a graph is important, particularly because this characteristic typically arises
in problems related to network reliability \cite{PrBa83}. Okamoto et al. \cite{OkUnUe05} proposed
$O(n + m)$ time algorithms for counting the numbers of independent sets and maximum independent
sets in a chordal graph. Lin et al. \cite{Li07,LiCh08} considered interval graphs and obtained
efficient linear $O (n)$ algorithms for counting the number of VCs, minimal VCs, minimum VCs, and
maximum minimal VCs in an interval graph. Recently, Lin and Chen \cite{LiCh09} presented $O (n^2)$
algorithms for the same vertex cover properties in a trapezoid graph.

In graph theory, the notions of vertex cover and independent set are dual to each other. A subset
$I \subseteq V$ is called an independent set of $G$ if and only if every edge in $E$ is incident on
no more than one vertex in $I$.

\begin{thm}[\cite{CoLeRiSt01}]
\label{thm-ind} A set $C$ is a minimal (minimum) vertex cover of $G$ if and only if $V - C$ is an
maximal (maximum) independent set of $G$.
\end{thm}

A matching in a graph is a set of edges in which no two edges are adjacent. A single edge in a
graph is obviously a matching. A maximum matching is a matching with the maximum cardinality and
the cardinality of the maximum matching is called a matching number. Currently, the best known
algorithm for the constructing maximum matching in general graphs is Edmonds' algorithm
\cite{Ga80,MiVa80}, which is based on the alternating paths, blossom and shrinking and runs in time
$O(|E| \sqrt{ |V|})$. For special classes of graphs, such as interval, chordal and permutation
graphs, more efficient algorithms were designed (see \cite{AnAtChLe00,Ch96,ChPaCh97} and references
therein). Ghosh and Pal \cite{GhPa05} presented an efficient algorithm to find the maximum matching
in trapezoid graphs, which turns out to be not correct.

The rest of the paper is organized as follows. In Section 2 we introduce the binary indexed tree
data structure. In Section 3 we design $O (n \log n)$ time dynamic programming algorithm for
calculating the size of the minimum vertex cover, the number of vertex covers and the number of
minimum vertex covers, improving the algorithms from \cite{LiCh09}. In Section 4 we present family
of counterexamples for $O (n^2)$ algorithm from \cite{GhPa05} for finding the maximum matching in
trapezoid graphs.

\section{Binary indexed data structure}

The binary indexed tree (BIT) is an efficient data structure introduced by Fenwick \cite{Fe94} for
maintaining the cumulative frequencies. The BIT was first used to support dynamic arithmetic data
compression and algorithm coding.

Let $A$ be an array of $n$ elements. The binary indexed tree supports the following basic
operations:
\begin{enumerate}[($i$)]
\item for given value $x$ and index $i$, add $x$ to the element $A (i)$, $1 \leq i \leq n$;
\item for given interval $[1, i]$, find the sum of values $A (1), A (2), \ldots, A (i)$, $1 \leq i \leq n$.
\end{enumerate}


Naive implementation of these operations have complexities $O (1)$ and $O(n)$, respectively. We can
achieve better complexity, if we speed up the second operation which will also affect the first
operation. Another approach is to maintain all partial sums from $A(1)$ to $A(i)$ for $1 \leq i
\leq n$. This way the operations have complexities $O (n)$ and $O (1)$, respectively.

The main idea of binary indexed tree structure is that sum of elements from the segment $[1, i]$
can be represented as sum of appropriate set of subsegments. The BIT structure is based on
decomposition of the cumulative sums into segments and the operations to access this data structure
are based on the binary representation of the index. This way the time complexity for both
operations will be the same $O(\log n)$. We want to construct these subsegments such that each
element is contained in at most $\log n$ subsegments, which means that we have to change the
partial sums of at most $\log n$ subsegments for the first procedure. On the other hand, we want to
construct these subsegments such that for every $1 \leq i \leq n$ the subsegment $[1, i]$ is
divided in at most $\log n$ subsegments. It follows that these subsegments can be $[i - 2^{r(i)} +
1, i]$, where $r(i)$ is the position of the last digit $1$ (from left to right) in the binary
representation of the index $i$. We store the sums of subsegments in the array $Tree$ (see
Algorithm 1 and Algorithm 2), and the element $Tree (i)$ represents the sum of elements from index
$i - 2^{r(i)} + 1$ to $i$. The structure is space-efficient in the sense that it needs the same
amount of storage as just a simple array of $n$ elements. Instead of sum, one can use any
distributive function (such as maximum, minimum, product). We have the following

\begin{prop}
Calculating the sum of the elements from $A (1)$ to $A (i)$, $1 \leq i \leq n$, and updating the
element $A (i)$ in the binary indexed tree is performed in $O (\log n)$ time.
\end{prop}

The fundamental operation involves calculating a new index by stripping the least significant 1
from the old index, and repeating this operation until the index is zero. For example, to read the
cumulative sum $A (1) + A (2) + \ldots + A (11)$, we form a sum
$$
Tree (11) + Tree (10) + Tree (8) = \Big(A (11) \Big) + \Big (A (9) + A (10) \Big) + \Big(A(1) + A
(2) + \ldots + A (8)\Big ).
$$
An illustrative example is presented in Table 1.

\begin{table}[ht]
\centering 
\begin{tabular} {l|llllllllllllllll}
\toprule

$i$ & 1 &  2 &  3 &  4  & 5 &  6 &  7 &  8 &  9 &  10 & 11 & 12 & 13 & 14 & 15  & 16 \\
\midrule
$A$ &  1 &  0 &  2 &  1 &  1 &  3 &  0 &  4 &  2 & 5 &  2 &  2 &  3 &  1 &  0 &  2 \\
$Tree$ &  1 & 1 &  2 & 4 &  1 &  4 & 0 & 12 & 2 & 7 & 2 &  11 & 3 &  4 & 0 & 29\\
Cumulative sums &  1 &  1 &  3 &  4  & 5 &  8 &  8 &  12 & 14 & 19 & 21 & 23 & 26 & 27 & 27 & 29 \\
\bottomrule
\end{tabular}

\caption{An example with array $A$ of length $n = 16$, binary indexed tree $Tree$ and cumulative
sums.}
\end{table}


Using the fast bitwise AND operator, we can calculate the largest power of 2 dividing the integer
$n$ simply as $r (n) = n \ AND \ (-n)$.

\begin{algorithm}
    \KwIn{The value $value$ and element index $index$.}

    \While {$index \leq n$}
    {
        $Tree [index] = Tree [index] + value$\;
        $index = index + (index \ \mathbf{and} \ (-index))$\;
    }
    \caption{ Updating the binary indexed tree. }
\end{algorithm}

\begin{algorithm}
    \KwIn{The index $index$.}
    \KwOut{The sum of elements $A[1], A [2], \ldots, A [index]$.}

    $sum = 0$\;
    \While{$index > 0$}
    {
        $sum = sum + Tree[index]$\;
        $index = index - (index \ \mathbf{and} \ (-index))$\;
    }
    \Return $sum$\;

    \caption{ Calculating the cumulative sum. }
\end{algorithm}

\section{The algorithm for minimum vertex covers}

Let $T = \{1, 2, \ldots, n\}$ denote the set of trapezoids in the trapezoid graph $G = (V, E)$. For
simplicity, the trapezoid in $T$ that corresponds to vertex $i$ in $V$ is called trapezoid $T (i)$.
Without loss of generality, the points on each horizontal line of the trapezoid diagram are labeled
with distinct integers between $1$ and $2n$.

For simplicity and easier implementation, we will consider the maximum independent sets (according
to Theorem \ref{thm-ind}). At the beginning, we add two dummy trapezoids $0$ and $n + 1$ to $T$,
where $a(0) = b(0) = c(0) = d(0) = 0$ for the vertex 0 and $a(n+1) =b(n+1) = c(n+1) =d(n+1) = 2n+1$
for the vertex $n + 1$. Trapezoid $i$ lies entirely to the left of trapezoid $j$, denoted by $i \ll
j$, if $b(i) < a(j)$ and $d(i) < c( j)$. It follows that $\ll$ is a partial order over the
trapezoid set $T$ and $(T ,\ll)$ is a strictly partially ordered set.

In order to get $O (n \log n)$ algorithms, we will use binary indexed tree data structure for sum
and maximum functions.

\subsection{The size of maximum independent set and the total number of independent sets}

Arrange all trapezoids by the upper left corner $a (i)$. Let $max\_ind (i)$ denote the size of the
maximum independent set of $T(1), T(2), \ldots, T(i-1), T (i)$ that contain trapezoid $T(i)$. For
the starting value, we set $max\_ind (0) = 0$. The following recurrent relation holds
$$
max\_ind (i) = \max \{1 + max\_ind (k), 0 \leq k < i \mbox{ and } T (k) \ll T (i) \}.
$$
This produces a simple $O (n^2)$ dynamic programming algorithm (similar to \cite{LiCh09}).


To speed up the above algorithm, we will store the values from $max\_ind$ in the binary indexed
tree $cum\_max$ for maintaining the partial maximums. First initialize each element of $cum\_max$
with $-1$. We need an additional array $index$, such that $index (j)$ contains the index of the
trapezoid with left or right coordinate equal to $j$. We are going to traverse the coordinates from
$0$ to $2n + 1$, and let $j$ be the current coordinate. In order to detect which already processed
trapezoids are to the left of the current trapezoid $T (i)$, we calculate the value $max\_ind (i)$
when the current coordinate is the left upper coordinate of $a (i)$ and insert it in the binary
indexed tree when it is the right upper coordinate $b (i)$. Then we have two cases:

\begin{enumerate}[($i$)]
\item coordinate $j$ is the upper left coordinate of the trapezoid $T (i)$.
We only need to consider the trapezoids that have lower right coordinate smaller than $c (i)$. As
explained, these trapezoids have their corresponding values stored in the segment $[1, c(i)]$ in
the binary indexed tree and therefore we can calculate $max\_ind (i)$ based on the maximum values
among $cum\_max(1), cum\_max(2), \ldots, cum\_max(c(i))$.

\item coordinate $j$ is the upper right coordinate of the trapezoid $T (i)$. In this case,
the value $max\_ind (i)$ is already calculated and we put this value in the binary indexed tree at
the position $d (i)$.
\end{enumerate}

The final solution is $max\_ind(n+1) - 1$. The pseudo-code is presented in Algorithm 3.

\begin{algorithm}
    \KwIn{The trapezoids $T$ and the array $index$.}
    \KwOut{The number of maximum independent sets stored in the array $max\_ind$.}

    Initialize binary indexed tree for maximum $cum\_max$\;
    \For{$j = 0$ \KwTo $2n + 1$}
    {
        $i = index [j]$\;
        \If{$a [i] = j$}
        {
            $max\_ind [i] = \mathbf{Calculate} (c [i]) + 1$\;
        }
        \Else(\tcp*[h]{  $b [i] = j$})
        {
            $\mathbf{Update} (d [i], max\_ind [i])$;
        }
    }
    \Return $max\_ind [n + 1] - 1$\;

    \caption{ The size of the maximum independent set. }
\end{algorithm}

\medskip

The set of all VCs and the number of VCs of $G$ are denoted by $VC(G)$ and $|VC(G)|$, respectively.
Let $N_G(v)$ be the set of vertices adjacent to $v$ in the graph $G$. In \cite{LiCh09} the authors
proved the following result.

\begin{lemma}
\label{le-1} For a graph $G$ and arbitrary vertex $v \in V$, it holds
$$
|VC (G)| = |VC (G - v)| + |VC (G - v - N_G (v))|.
$$
\end{lemma}

For counting the total number of independent sets, we can use the same method as above. The only
difference is that we will have sum instead of maximum function in the binary indexed tree. Let
$num\_ind (i)$ denote the number of independent sets of trapezoids $T(1), T(2), \ldots, T(i-1), T
(i)$ that contain trapezoid $T(i)$. According to Lemma \ref{le-1} the following recurrent relation
holds
$$
num\_ind (i) = 1 + \sum_{0 \leq k \leq i, \ T (k) \ll T (i)} num\_ind (k).
$$
The total number of independent sets is simply $num\_ind (n + 1) - 1$ and time complexity is the
same $O (n \log n)$.

\subsection{The number of maximum independent sets}

For counting the number of maximum independent sets, we will need one additional array
$num\_max\_ind$, such that $num\_max\_ind (i)$ represents the number of maximum independent sets
among trapezoids $T (1), T (2), \ldots, T (i- 1), T (i)$ such that $T (i)$ belongs to the
independent set. After running the algorithm for calculating the array $max\_ind$, for each $1 \leq
i \leq n$ we need to sum the number of independent sets of all indices $j$, such that $T (j) \ll T
(i)$ and $max\_ind (j) + 1 = max\_ind (i)$ in order to get the number of independent sets with
maximum cardinality. The pseudo-code of the algorithm for counting the number of maximum
independent sets from \cite{LiCh09} is presented in Algorithm 4.

\begin{algorithm}
    \KwIn{The trapezoids $T$ and the array $max\_ind$.}
    \KwOut{The number of maximum independent sets stored in the array $num\_max\_ind$.}

    $num\_max\_ind [0] = 1$\;
    \For{$i = 1$ \KwTo $n + 1$}
    {
        $num\_max\_ind [i] = 0$\;
        \For{$j = 0$ \KwTo $i - 1$}
        {
            \If{$(max\_ind [j] + 1 = max\_ind [i])$ \textbf{and} $(a [j] > b [i])$ \textbf{and} $(c [j] > d [i])$}
            {
                $num\_max\_ind [i] = num\_max\_ind [i] + num\_max\_ind [j]$\;
            }
        }
    }
    \Return $num\_max\_ind [n + 1] - 1$\;

    \caption{ The number of maximum independent sets. }
\end{algorithm}

We will use again binary indexed trees for designing $O (n \log n)$ algorithm. The main problem is
how to manipulate the partial sums. Namely, for each trapezoid $T(i)$, we need to sum the values
$num\_max\_ind (j)$ of those trapezoids $T(j)$ which are entirely on the left of $T(i)$ with
additional condition $max\_ind (j) + 1 = max\_ind (i)$.

This can be done by considering all pairs $(k, k + 1)$ of two neighboring values $1 \leq k \leq
max\_ind (n + 1)$. Using two additional arrays for the coordinates and trapezoid indices, we can
traverse trapezoids $T (i)$ from left to right, such that $max\_ind (i) = k$ or $max\_ind (i) = k +
1$. First, for each $1 \leq k \leq max\_ind (n + 1)$ one needs to carefully preprocess all
trapezoids and store the indices and coordinates of all trapezoids $T (i)$ with $max\_ind (i) = k$
in order to keep the memory limit $O (n)$. Therefore, all trapezoids will be stored in the dynamic
array of arrays $S$, such that $S (k)$ contains all trapezoids $T (i)$ with $max\_ind (i) = k$.
Then we fill the binary indexed tree with the values for trapezoids with $max\_ind (i) = k$, and
calculate the value $num\_max\_ind (i)$ for trapezoids with $max\_ind (i) = k + 1$, as before.
Instead of resetting the binary indexed tree for each $k$, we can again traverse the trapezoids
with $max\_ind (i) = k$ and update the values of BIT to $0$.

Since every trapezoid will be added and removed exactly once from the binary indexed tree data
structure, the total time complexity is $O (n \log n)$. This reusing of the data structure is novel
approach to the best of our knowledge and makes this problem very interesting.
\medskip

We conclude this section by summing the results in the following theorem.

\begin{thm}
The proposed algorithms calculate the size and the number of maximum independent sets, as well as
the total number of independent sets in time $O (n \log n)$ of trapezoid graph $G$ with $n$
vertices.
\end{thm}

One can easily extend this algorithm for efficient construction the independence polynomial of a
trapezoid graph
$$
I (G, x) = \sum_{k = 0}^{\alpha (G)} s_k x^k,
$$
where $s_k$ is the number of independent sets of cardinality $k$ and $\alpha(G)$ is the
independence number of $G$.

\section{The algorithm for the maximum matching}

Ghosh and Pal \cite{GhPa05} presented an efficient algorithm for finding the maximum matching in
trapezoid graphs. The proposed algorithm takes $O(n^2)$ time and $O(n + m)$ space. They defined the
right spread $f(i)$ of a trapezoid $T (i)$ as the maximum $\max \{b (i), d (i)\}$. Let $M$ be the
set of edges which form a matching of the graph $G$. Deletion of a vertex from the graph means
deletion of that vertex and its adjacent edges.

The algorithm is designed as follows. Calculate the right spread $f (i)$ and arrange all trapezoids
by $f$ in ascending order for $1 \leq i \leq n$ in linear time $O (n)$. Select the trapezoid $i$
with the minimum value of the right spread $f (i)$. After that, select the second minimum value $f
(j)$. If trapezoids $T (i)$ and $T (j)$ are adjacent, then put the edge $(i, j)$ in the maximum
matching $M$, remove the vertices $i$ and $j$ from the graph $G$ and mark the corresponding
elements of the array $f$. If $(i, j) \not \in E (G)$ then continue selecting the next minimum
elements $f (j')$ from the array $f$ until the trapezoids $i$ and $j'$ are adjacent. In that case
remove the vertices $i$ and $j'$, put the edge $(i, j')$ in the maximum matching $M$ and mark the
corresponding elements of $f$; otherwise remove the trapezoid $i$ from the graph $G$ since it
cannot be matched. Continue this procedure as long as $|V (G)| > 1$. Since for each trapezoid we
potentially traverse all trapezoids with larger values $f (i)$, the time complexity of the proposed
algorithm is $O (n^2)$.

\begin{figure}[ht]
  \center
  \includegraphics [width = 13cm]{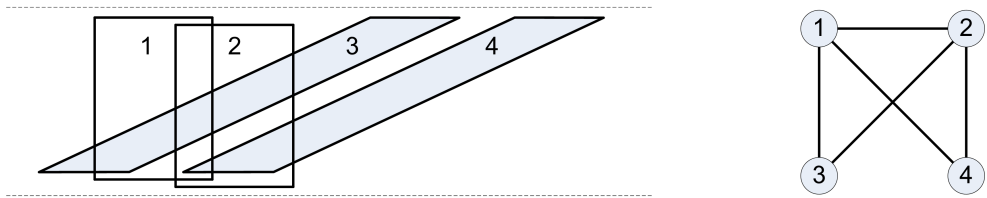}
  \caption { The minimal counterexample for the matching algorithm. }
\end{figure}

Obviously, the algorithm works for any set of $n = 3$ trapezoids . In the configuration on Figure
2, there are four trapezoids and only trapezoids $T (3)$ and $T (4)$ are not adjacent. The
trapezoids are ordered such that $f (1) < f (2) < f (3) < f (4)$. By proposed algorithm in
\cite{GhPa05}, we will match the trapezoids $1$ and $2$ and get the maximal matching of cardinality
1. But obviously, the size of the maximum matching is 2 by pairing the trapezoids $(1, 3)$ and $(2,
4)$. Therefore, this algorithm is not correct and gives only an approximation for the maximum
matching. One can easily construct a family of counterexamples based on this minimal example by
adding trapezoids to the left and right.

We leave as an open problem to design more efficient algorithm for finding maximum matching in
trapezoid graphs.

\vspace{0.4cm} {\bf Acknowledgement. }  This work was supported by Research Grants 174010 and
174033 of Serbian Ministry of Education and Science.

\end{document}